\documentclass[12pt,preprintnumbers,notitlepage,tightenlines]{revtex4-1} 
\usepackage{amsmath}  % needed for \tfrac, \bmatrix, etc.
\usepackage{amsfonts} % needed for bold Greek, Fraktur, and blackboard bold
\usepackage{graphicx} % needed for figures

\renewcommand\Im{\operatorname{Im}}

\global\long\def\dd{\mathrm{d}}
\def\I{\mathcal{I}}
\begin{document}

\preprint{Alberta Thy 5-14}
\title{Current distribution in an infinite plate}

\author{Andrzej Czarnecki}
\affiliation{Department of Physics, University of Alberta, Edmonton, Alberta,
Canada T6G 2E1}

\begin{abstract}
Distribution of the electric potential in a
very long plate (for example a long metal ruler) is
determined. This is achieved by conformally mapping the plate into a
plane, simplifying the geometry of boundary conditions.  Singularities
of the potential are discussed as well as their regularization by the
final size of electrical contacts.  An analogy with renormalization is
pointed out.   Results are compared with previous studies.
\end{abstract}
\maketitle

\section{Introduction}
\begin{figure}[h!]
\centering
\includegraphics[width=5in]{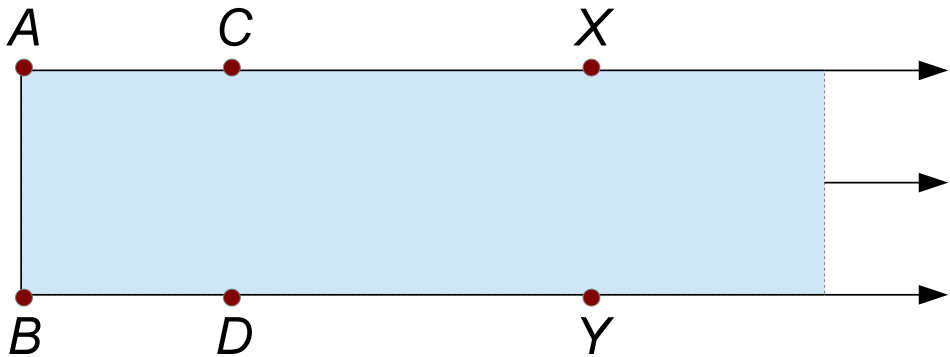}
\caption{A semi-infinite metal ruler.  A  voltage difference of 1 V is
  applied to points $A$ and $B$, and as a result 0.1 V is measured
  between $C$ and $D$.  What is the voltage difference between $X$ and
$Y$ as a function of their distance from the  edge $AB$?}
\label{fig:ruler}
\end{figure}

This paper is motivated by an intriguing question posed in the
excellent collection of physics problems
\cite{guideToPhysics1}, based on an exam problem from the prestigious
Moscow Institute of Physics and Technology ``Phys-Tech''.  Imagine a
long ruler made of metal, shown in Fig.~\ref{fig:ruler}.  We apply voltage
$ U_s=1$ V to its corners $A$ and $B$, and 1 cm away from the edge
we find the voltage difference between points $C$ and $D$ to be
$U_d = 0.1$ V
($AC=1$ cm).  The question is how the voltage difference behaves
further down the ruler.  The thickness of the ruler is neglected so
the problem is two-dimensional.

It seems we can reproduce this situation in reality.  However, note
that the real current leads always have a finite extension area (or,
in the idealized 2D situation, a finite length) of contact and are never
point-like.  In fact, we will see below that the original problem
cannot be solved.  However, it is a very inspiring problem that can be
reformulated and used to illustrate the important mathematical
technique of conformal mapping
\cite{jackson:107,furman:1134,mittag:207} in an intuitive setting.  It
can also be used to illustrate the concept of renormalization
\cite{olness:306,delamotte:170,mead:935}, where microscopic details
  of the system are not known but predictions of relationships among
  macroscopic measurements are possible.

\section{Electric potential distribution}
The electric potential  $U$ and the density of current $j$ flowing in the ruler
are related by Ohm's law: the current flows along the gradient of the
potential (in an isotropic medium) and the proportionality coefficient
is the  conductivity $\sigma$,
\begin{equation}
\vec j = -\sigma \vec \nabla U.
\end{equation}
Away from the contact points $A$ and $B$, the current is conserved,
$\vec  \nabla \cdot \vec j = 0$, so, inside the ruler, the potential satisfies the
Laplace equation,
\begin{equation}
\nabla^2 U = 0.
\end{equation}
What is the behavior of $U$ near the contact point $A$? Consider a
small circle around $A$ with the radius $\epsilon$ much less than the
width of the ruler.  Since the current is conserved, its density is
inversely proportional to the radius of the circle, $j\sim 
1/\epsilon$.  We see that $j$ has a pole at $A$ (as well as at $B$),
and the potential has logarithmic singularities at those points.  For
this reason the problem cannot be solved as originally formulated: if
the potential difference at the source $U_s$ is finite, the current
cannot flow and is zero everywhere, and so is the potential difference
between any points further along the ruler, like $C$ and $D$.  If, on
the other hand, the potential difference $U_d$ at some distance from
the source is non-zero, the source voltage $U_s$ must be infinite.

In the remainder of this section we will find the potential
distribution inside the ruler, assuming infinite $U_s$.  In the
following section we will use this result to reformulate the problem
and make it solvable.   

The current does not flow across the edges of the ruler, so along the
edges the component of the current density perpendicular to the edge
vanishes.  This is the boundary condition that $U$ must satisfy: the
gradient of $U$ must not have components perpendicular to the edges.
This boundary problem is much easier to solve a simpler geometry, so
we use a conformal mapping to map the ruler into an infinite
half-plane.

We introduce a complex variable $z=x+iy$ as a coordinate in the ruler:
point $B$ (see Fig.~\ref{fig:ruler}) is at the origin $z_B=0$ and the
scale is chosen such that point
$A$ is at $z_A = i$.  If we consider any (twice differentiable)
function of $z$, its real and imaginary components automatically
fulfill the Laplace equation (see the Appendix). 

We now introduce a new complex variable $w = \exp\pi z$.  An important
property of such complex mapping is that it preserves angles (see the
Appendix).  If we find a potential $U(w)$ whose
equipotential lines are perpendicular to the images of the ruler
boundaries in the mapping $w(z)$, the function $U\left( w(z) \right)$
will be automatically the correct solution of our problem in the more
complicated geometry of the ruler itself.

\begin{figure}[h!]
\centering
\includegraphics[width=5in]{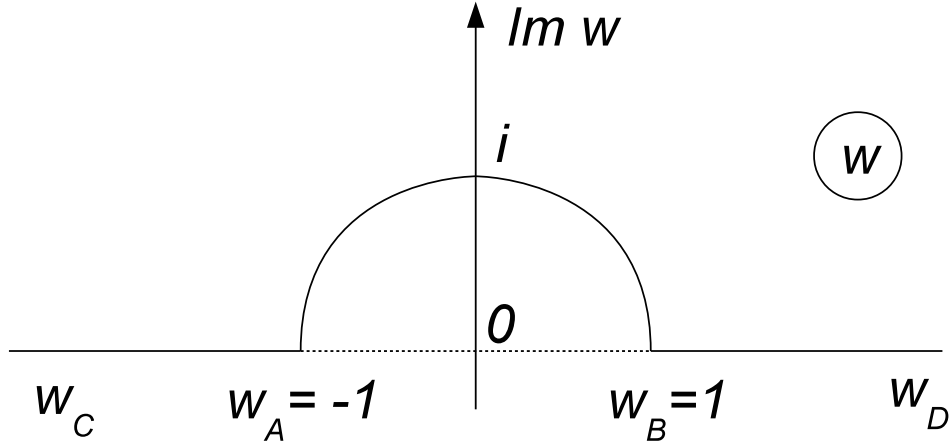}
\caption{Image of the ruler in the conformal mapping $w(z)=\exp\pi
  z$.  If the ruler extended to infinity both to the right and to the left, its
  image would be the whole upper half-plane Im$(w)\leq 0$.  The
  additional, left-hand side half of the ruler would all be mapped onto the
  unit semicircle  around the origin.  The curved edge of the
  semicircle is the image of the edge $AB$ of the ruler.  The imaginary
  $u$ axis is the image of the horizontal symmetry axis of the ruler,
  Im$(z) = 1/2$.}
\label{fig:rulerImage}
\end{figure}

The image of the ruler under $z\to w(z) = \exp\pi z$ is shown if
Fig.~\ref{fig:rulerImage}: the current sources $A$ and $B$ are mapped onto
two points on the $x_w$ axis, $w_A = (-1,0)$ and $w_B = (1,0)$.  

The long edges of the ruler are mapped onto the infinite parts of the
$x_w$ axis: the upper edge onto the negative part $x_w \leq -1$ and
the lower one onto the positive part $x_w \geq 1$.  We now see the
advantage of using the map $w(z)$:  since the sources lie on the $w_x$
axis, any function that depends only on the distances from these
sources is symmetric with respect to a reflection in that axis, and
lines of its constant real or imaginary part can intersect that axis
only at right angles.

The horizontal line dividing
the original plate in half ($\Im z=1/2$) is mapped onto the positive imaginary semi-axis.
$w=0$ corresponds to $z=-\infty$. 

The solution of the Laplace equation is \begin{equation}
U\left(w=x_{w}+iy_{w}\right)=C\ln\left(\frac{\mbox{distance from
    }w_{A}}
{\mbox{distance from
  }w_{B}}\right)^{2}=C\ln\frac{\left(x_{w}+1\right)^{2}+y_{w}^{2}}{\left(x_{w}-1\right)^{2}+y_{w}^{2}}.
\label{semiInfPlate:potInW}\end{equation}
The overall normalization $C$ depends on the amount of current 
flowing into the ruler and the conductivity.  We will discuss its
determination below.

As we have already noticed, this solution, by symmetry,
satisfies the boundary condition that the gradient of the potential
at the boundaries of the semi-infinite plate should have no component
perpendicular to the boundary (since no current flows through the
boundary). In the $w$ plane this is satisfied by \eqref{semiInfPlate:potInW}.
Namely, equipotential lines form some closed figures around $w=1$
and $-1$, symmetric with respect to $y_{w}\to-y_{w}$. This shows
that they are orthogonal to the upper and lower plates. (It is less
obvious that they are also orthogonal to the vertical edge of the
plate which is mapped onto a semicircle connecting $w_{A}$ with $w_{B}$.
But they must be, by symmetry (in the $z$-plane) between the right and
the left halves of the infinite plate $\Im z\le 1$.)

All we now need is to express $x_{w}$ and $y_{w}$ in
eq.~(\ref{semiInfPlate:potInW}) by the original $x,y$,
\begin{eqnarray}
x_{w} & = & \exp\pi x\cos\pi y\\
y_{w} & = & \exp\pi x\sin\pi y.\end{eqnarray}
We find \begin{equation}
U\left(x,y\right)=C\ln\frac{1+2e^{-\pi x}\cos\pi y+e^{-2\pi x}}{1-2e^{-\pi x}\cos\pi y+e^{-2\pi x}}\end{equation}
and the voltage difference between the upper and the lower edge ($y=1$
and $0$) is\begin{equation}
U_x =4C\ln\frac{1+e^{-\pi x}}{1-e^{-\pi
    x}}\to8Ce^{-\pi x}\quad\mbox{for }x\gg 1/\pi.\end{equation} 
We see that far from the edge the voltage drops exponentially
and that the exponent does not depend on the resistivity. What sets
the scale is the width of the plate (taken here as the unit of length). 

\subsection{Current density and the normalization}

We now express the normalization constant $C$ in terms of the current
flowing into the ruler. To this end, consider the potential near the
origin $x=y=0$, 
\begin{equation}
U\left(x\to 0,y\to 0\right)=2C\ln\frac{2}{\pi r},\quad r\equiv
\sqrt{x^2+y^2}.
\label{eq:corner}
\end{equation}
The resulting current density is
\begin{equation}
\vec j = {2C\sigma \hat r\over r},
\end{equation}
relating the normalization $C$ to the total current $\I$ flowing through
the ruler, 
\begin{equation}
C={\I\over \pi \sigma}.
\end{equation}
\subsection{A model of electrical contacts}
\begin{figure}[h!]
\centering
\includegraphics[width=3in]{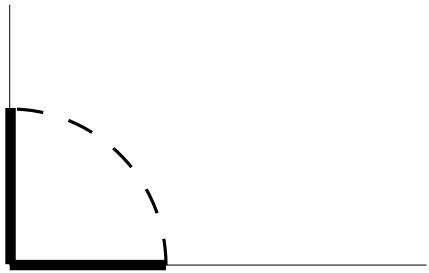}
\vspace*{-15mm}
\caption{A model of a finite contact supplying current to the ruler.
  The contact is shown with a thick line.  The dashed line shows an
  equipotential line (in case of point-like contacts) of similar length
as the real contact.}
\label{fig:contact}
\end{figure}

Suppose the current is supplied to a ruler by contacts that wrap
around the corners of the ruler near points $A$ and $B$ in
Fig.~\ref{fig:ruler}.  We assume the length of the contacts to be a
small fraction $s$ of the width $AB$, $s\ll 1$.   In order to find the
resistance between such finite contacts, we divide the ruler into a
tiny part between the contact and a quarter-circle of similar size (an
equipotential line in case of point-like contacts), as shown in
Fig.~\ref{fig:contact}.  
 
The resistance between the contacts is dominated by the bulk of the
ruler between the quarter-circles at both corners; for $s\ll 1$, it
behaves like $\ln{1\over s}$ (see Eq.~(\ref{eq:corner})).    The
quarter-circles themselves have all dimensions on the order of $s$, so
their resistance adds only a constant term to that logarithm and can
be treated as a subleading correction.    

We thus see that the voltage $U_s$  applied to the corners 
is related to the current flowing through the ruler by
\begin{equation}
U_s = {4\I \over \pi \sigma}(c-\ln s),
\end{equation}
where, in the limit $s\ll 1$,  $c$ is a constant dependent only on the
detailed geometry of the contacts but not their size.  

It is convenient to express the normalization constant of the
potential distribution by the applied voltage,
\begin{equation}
C =\frac{U_s}{4 (c-\ln s)},
\end{equation}
so that the voltage difference between points on the upper and lower
edge of the ruler at the distance $x$ from the left edge is
\begin{equation} 
U_x =\frac{U_s}{c-\ln s}\ln\frac{1+e^{-\pi x}}{1-e^{-\pi  x}}.
\label{eq:full}
\end{equation} 
For simplicity, from now on, we assume that the distances from the
left edge of the ruler are sufficiently large for the simple asymptotic
formula  to be valid (this is true for $x$ much larger than about a
third of the ruler width),
\begin{equation} 
U_x \to   \frac{2U_s}{c-\ln s}e^{-\pi x}\quad\mbox{for }x\gg 1/\pi.
\label{eq:asymp}
\end{equation}
We now want to use the voltage difference 
measured at some distance $x=d$ ($d$ equals to the dimensionless ratio
$AC/AB$ in Fig.~\ref{fig:ruler}),
\begin{equation}
  \label{eq:8}
  U_d = \frac{2U_s}{c-\ln s}e^{-\pi d},
\end{equation}
to express the voltage difference at another point at a dimensionless
distance $x$ from the edge,
\begin{equation}
  \label{eq:1}
U_x = U_d e^{-\pi(x-d)}.  
\end{equation}
We see that just knowing the source voltage $U_s$ and one voltage $U_d$ at
some distance from the edge is not sufficient to determine it at other
distances.  The reason is that there are three unknowns: the
width of the ruler $W$, the size of the contact, and the voltage distribution.
$U_s$ has the character of an ``unrenormalized'' quantity whose
relationship to physical quantities like $U_x$ depends on the
microscopic details of the contacts, parameterized by $s$ and $c$.

However, if we instead know two voltages measured at some distances
away from the edge, say $D_{1,2} =d_{1,2} W$, we can find the
voltage distribution everywhere (away from the origin).  This is
because the voltages away from the source depend on one universal
combination of the applied voltage and the details of contacts.  
A new information has been added without increasing the number of
given parameters. This leads us to the new formulation of the problem.

\section{How can this problem be  formulated}
If the measured voltage is given at two pairs of points along the ruler, like
$C,D$ and $X,Y$ in Fig.~\ref{fig:ruler}, it can be found everywhere else.  We use the notation $D_1 =
AC$, $D_2=AX$, $W=AB$; and the distances are expressed in terms of the
width, $d_{1,2} = D_{1,2}/W$.  We then have  
\begin{eqnarray}
  \label{eq:2}
  U_{d_{1,2}} &=& \frac{U_s}{c-\ln s} e^{-\pi d_{1,2}},
\nonumber \\
  U_x &=& \frac{U_s}{c-\ln s} e^{-\pi x} = U_{d_1} e^{-\pi (x-d_1)}.
\end{eqnarray}
It remains to eliminate the unknown width of the ruler using the two
given voltages,
\begin{equation}
  \label{eq:3}
  W=\pi \frac{D_2 -D_1}{\ln \frac{U_{d_1}}{U_{d_2}}},
\end{equation}
and finally
\begin{equation}
  \label{eq:4}
  U_x = U_{d_1} \left(  \frac{U_{d_2}}{U_{d_1}} \right)^\frac{x-d_1}{d_2-d_1}.
\end{equation}
We have been able to eliminate the unknown microscopic parameters of
the problem in terms of the measured macroscopic quantities.  It seems
that an analytical solution (\ref{eq:4}) in terms of $U_{d_{1,2}}$ is only possible
if the asymptotic formula (\ref{eq:asymp}) is valid for $d_{1,2}$.
Otherwise the transcendental equations resulting from (\ref{eq:full})
have to be solved numerically or geometrically.

\section{Discussion of another approach}
We now want to compare our study with the solution proposed in
\cite{guideToPhysics1}.  The starting point there is a comparison of
potential differences $U_x$ at
two nearby pairs of points along the ruler, 
\begin{equation}
  \label{eq:5}
  U_{x+dx} - U_x = \alpha(x) U_x \dd x,
\end{equation}
which is certainly true as long as the coefficient $\alpha(x)$ is a
function of $x$.  It is then argued that since the ruler is
semi-infinite, $\alpha(x) = \alpha$ is a constant independent of the
position (as if all points had the same properties).  This seems to be
wrong, since the ruler is only {\em semi}-infinite, so points closer
to the left edge are really different from points further down.
However, the differential equation thus obtained,
\begin{equation}
  \label{eq:6}
  U_x^\prime = \alpha U_x,
\end{equation} 
is also satisfied by our asymptotic solution (\ref{eq:asymp}).  This
is because for $x\gg 1$ (points far away from the beginning of the
ruler), the dependence of $\alpha(x)$ on $x$ becomes very weak.  

If one treats the equation (\ref{eq:6}) as valid exactly
everywhere, one gets the (incorrect) solution 
\begin{equation}
  \label{eq:7}
  U_x = U_s \left( \frac{U_s}{U_d} \right)^{-x/d}
\end{equation}
But is this really incorrect?  After all, it seems to agree with
(\ref{eq:4}) if we set there $d_1\to 0$.  The point is that
(\ref{eq:4}) is {\em not valid} unless $d_1 \gg 1$; it was derived
using the asymptotic formula (\ref{eq:asymp}).  So, one cannot study
the limit  $d_1\to 0$ with that equation.  

\subsection{Numerical example} 
Consider two rulers, one with the width $W_1 =30$ mm, another with $W_2 = 13$
mm, as are typically used in classroom and in the mechanical shop.  We
use Eq.~(\ref{eq:full}) to
adjust contact parameters, $c-\ln s$, to ensure that 1~V supplied
to points A and B (see Fig.~\ref{fig:ruler}) results in 0.1~V measured 1
cm away from the edge.  We find $c_1-\ln s_1 = 7.3$,  $c_2-\ln s_2 =
1.8$.  

Now, using again  Eq.~(\ref{eq:full}), we can determine the voltage
difference measured at any distance away from the left edge.  Choose
the distance equal
20 mm, that is twice the distance where 0.1~V was measured.  In the
broader ruler we find $U_1(20\ \mathrm{ mm}) = 0.03$ V, and in the
narrower $U_2(20\ \mathrm{ mm}) = 0.009$ V. 

For comparison, the incorrect formula (\ref{eq:7}) gives in both
cases 0.01 V, independent of the width. 
It is accidental that this result is numerically close to that for the
13 mm ruler.  For an even narrower one, with $W_3=10$ mm,
Eq.~(\ref{eq:7})  still gives 0.01 V, while the correct result is
0.004 V.

\section{Summary} 
Somewhat surprisingly the problem of finding the voltage distribution
in a long ruler has turned out to involve some interesting mathematics
and physics.  We have seen that the solution can be found using a
conformal mapping.  The procedure is so simple and intuitive that it
can be used as a pedagogical illustration of this powerful method of
complex analysis.  The exact solution with point-like contacts has
logarithmic singularities.  These are regularized when finite size of
contacts is taken into account.  Finally, we have found that
relationships among measurements away from the contacts can be
expressed without reference to the microscopic details, in a way
analogous to the renormalization.

A solution for a very long ruler, to which
the voltage is applied in the middle, rather than at one end, is
trivially obtained from the one presented here.

\noindent
{\bf Acknowledgments:} I thank Alexander Penin and Mikhail Voloshin for helpful
discussions. This research was supported by Science and Engineering Research
Canada (NSERC).

\appendix*
\section{Conformal mapping: summary of properties}
We need two properties of the complex mapping $w=w(z)$ in the present analysis:
it is conformal, that is preserving angles at which curves intersect; and it preserves the
form of the Laplace equation.

\begin{figure}[h!]
\centering
\includegraphics[width=5in]{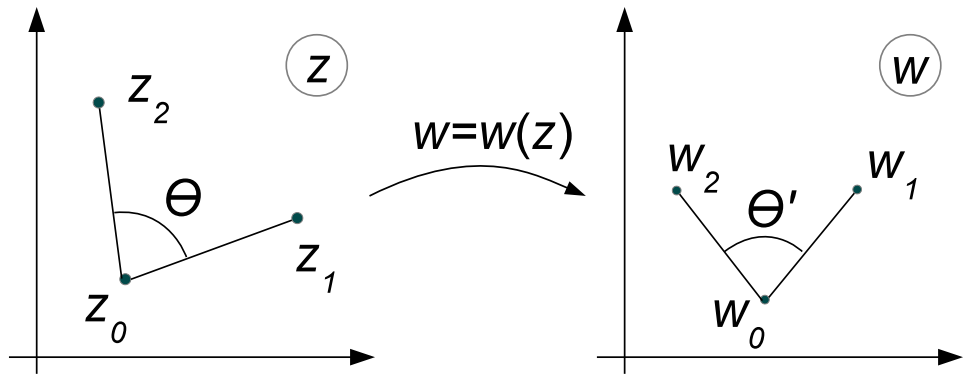}
% used epstopdf conformalAngle.eps in terminal
\caption{Conformal mapping preserves angles.}
\label{fig:conformalAngles}
\end{figure}

The first property follows from $w(z)$ being a differentiable
function; that is, its derivative $w'(z)$ does not depend on the
direction at which $z$ is approached.
To see this, consider the three points $z_{0,1,2}$ and
their images $w_{0,1,2}$, as  shown
in Fig.~\ref{fig:conformalAngles}.  We want to show that the two
angles $\theta$ and $\theta'$ are equal.  The points $z_{1,2}$ are
infinitesimally close to $z_0$, and for simplicity we
choose them to be at equal distance $r$ from $z_0$.  
The angle $\theta$ is just the difference of phase of $z_2-z_0\equiv
re^{i\phi_2}$ and $z_1-z_0\equiv re^{i\phi_1}$, 
\begin{equation}
\theta = \phi_2 -\phi_1 = {1\over i}\ln \frac{z_2-z_0}{z_1-z_0}.
\end{equation}
The angle $\theta'$ is found by expanding $w(z_{1,2})-w(z_0)$ in a Taylor
series; only one term of the expansion suffices since the points are infinitesimally close:
\begin{eqnarray}
\theta' &=&  {1\over i}\ln \frac{w(z_2) - w(z_0) }{w(z_1) - w(z_0)}
\nonumber \\
&=&  {1\over i}\ln \frac{w'(z_0)\left(z_2 - z_0\right) }{w'(z_0)\left(z_1 - z_0\right) }=\theta,
\end{eqnarray}
where we have used the independence of the derivative from the
direction. 

Regarding the Laplace equation, it is satisfied by any (twice-differentiable) function of a complex
variable $z=x+iy$,
\begin{equation}
\left(  {\dd^2 \over \dd x^2}  +   {\dd^2 \over \dd y^2} \right) f(z) =
\left[ \left( {\dd z \over \dd x} \right)^2
+  \left( {\dd z \over \dd y} \right)^2 \right]  
{\dd^2f \over \dd z^2}   = 0,
\end{equation}
since $\left( {\dd z \over \dd x} \right)^2
+  \left( {\dd z \over \dd y} \right)^2  = 1-1 =0$.  Any
twice-differentiable mapping $z\to w(z)$ preserves this property.

%\bibliographystyle{/Users/czar/Library/texmf/tex/latex/bibtexStyles/ac}
%\bibliography{/Users/czar/Documents/pro/Tables/Archive/phd}

\begin{thebibliography}{1}

\bibitem{guideToPhysics1}
S.~B. Cahn and B.~E. Nadgorny,
\newblock {\em A Guide to Physics Problems, Part 1: Mechanics, Relativity, and
  Electrodynamics} (Plenum, New York, 1994).

\bibitem{jackson:107}
J.~D. Jackson,
\newblock Am.~J.~Phys. {\bf 67}, 107 (1999).

\bibitem{furman:1134}
M.~A. Furman,
\newblock Am.~J.~Phys. {\bf 62}, 1134 (1994).

\bibitem{mittag:207}
L.~Mittag and M.~J. Stephen,
\newblock Am.~J.~Phys. {\bf 60}, 207 (1992).

\bibitem{olness:306}
F.~Olness and R.~Scalise,
\newblock Am.~J.~Phys. {\bf 79}, 306 (2011).

\bibitem{delamotte:170}
B.~Delamotte,
\newblock Am.~J.~Phys. {\bf 72}, 170 (2004).

\bibitem{mead:935}
L.~R. Mead and J.~Godines,
\newblock Am.~J.~Phys. {\bf 59}, 935 (1991).

\end{thebibliography}

\end{document}